\documentclass[preprint]{revtex4}
\pdfoutput=1
\usepackage{graphicx}
\usepackage[pdftex,colorlinks=true]{hyperref}
\usepackage{pdfpages}

\usepackage{mathtools}
\usepackage{bm}
\usepackage{multirow}

\def\ie{{\it i.e.}}
\def\eg{{\it e.g.}}

\def\C{{\rm C}}
\def\D{{\rm D}}
\def\G{{\rm G}}
\def\B{{\rm B}}

\def\R{{\rm R}}

\def\CG{{\rm C\rightarrow G}}
\def\CB{{\rm C\rightarrow B}}
\def\DB{{\rm D\rightarrow B}}
\def\DG{{\rm D\rightarrow G}}

\begin{document}

\title{Voting by Hands Promotes Institutionalised Monitoring in Indirect Reciprocity}

\author{Mitsuhiro~Nakamura}
\email{nakamuramh@soken.ac.jp}
\affiliation{%
  Department of Evolutionary Studies of Biosystems, \\
  SOKENDAI (The Graduate University for Advanced Studies), \\
  Hayama, Kanagawa 240-0193, Japan}

\author{Ulf Dieckmann}
\affiliation{%
  Evolution and Ecology Program, \\
  International Institute for Applied Systems Analysis, \\
  2361 Laxenburg, Austria}

\begin{abstract}
Indirect reciprocity based on reputation is a leading mechanism driving
human cooperation, where monitoring of behaviour and sharing reputation-related
information are crucial.
Because collecting information is costly, a tragedy of the commons can arise,
with some individuals free-riding on information supplied by others.
This can be overcome by organising monitors that aggregate information,
supported by fees from their information users.
We analyse a co-evolutionary model of individuals playing a social dilemma game and
monitors watching them; monitors provide information and players vote for a
more beneficial monitor.
We find that
(1) monitors that simply rate defection badly cannot stabilise
cooperation---they have to overlook defection against ill-reputed players;
(2) such overlooking monitors can stabilise cooperation if players vote for
monitors rather than to change their own strategy;
(3) STERN monitors, who rate cooperation with ill-reputed players badly,
stabilise cooperation more easily than MILD monitors, who do not do so;
(4) a STERN monitor wins if it competes with a MILD monitor; and
(5) STERN monitors require a high level of surveillance and achieve only
lower levels of cooperation, whereas MILD monitors achieve higher levels of
cooperation with loose and thus lower cost monitoring.
%
\end{abstract}
\keywords{%
  cooperation,
  indirect reciprocity,
  reputation,
  mutualism
}

\maketitle

\section{Introduction}
\label{sec:introduction}


The evolution of cooperation is a universal problem across
species~\cite{MaynardSmith1997,Axelrod1984,Nowak2006}.
To achieve cooperation, individuals often need to overcome a social dilemma: for
the population, all-out cooperation is the best, whereas for each individual, it
is better to free ride on the contributions of
others~\cite{Ostrom1990,Colman2006}.
Indirect reciprocity, among several other mechanisms, is a leading explanation
for the evolution of human
cooperation~\cite{Trivers1971,Alexander1987,Nowak1998a,Nowak2005,Sigmund2012}.
In indirect reciprocity, an individual helping another will be helped in the
future; cooperative individuals are highly valued and obtain help from others
because of their good reputation.


Indirect reciprocity fundamentally depends on the individuals' ability to
evaluate others and share information about their reputation (\eg, via gossip).
This requires an individual to obtain information about the others' reputation.
However, doing so is usually costly.
It demands considerable cognitive capacity to recognise and memorise others'
past actions~\cite{Milinski1998,Milinski2001,Suzuki2013}.
Gossip-based information sharing is vulnerable to liars who strategically spread
fake information~\cite{Nakamaru2004}.
As a recently emerging example, electronic marketplaces are adopting feedback
mechanisms to assess each seller.
However, customers often fail to submit such feedback as this involves
extra work~\cite{Gazzale2005,Gazzale2011,Masclet2012,Rockenbach2012}.
Consequently, the availability and reliability of information suffers from
a tragedy of the commons~\cite{Rockenbach2012,Rand2013}.


An important difference between a material good and information is
that information can be copied and distributed among many individuals at
negligible cost (even though its acquisition may be costly).
Therefore, as Arrow wrote, `it does not pay that everyone in a society
acquires this information, but only a number needed to supply the necessary
services'~\cite{Arrow2010}.
In human societies, such specialised servicing organisations gathering and
providing reputation information, \eg, modern credit companies and online
marketplaces, have played a major role~\cite{Fujiwara-Greve2012,Resnick2002}.
%
%
These organisations are maintained by their information users; the users demand
the supply of information and contribute fees in return.
This can be understood as a mutualism between monitoring services and
information users.
As far as we know, this has not been explored in the context of indirect
reciprocity.


In this study, we apply evolutionary game theory to the analysis of mutualism
between users of reputation-related information (\ie, the players) and
information-providing services (\ie, the monitors) in the context of indirect
reciprocity.
We present a co-evolutionary model in which players and monitors seek to adapt
their strategies through social learning.
The population of players is engaged in a social dilemma game called the
donation game;
from time to time, one player can decide whether to help another player or not.
The strategy can be unconditional: to always help, or to always refuse to help.
In this case, cooperation loses out.
But players can also use a conditional strategy, and help only those players who
have a good reputation.
We analyse whether competition between information providers can lead to
cooperation in the population of players.
%


In our evolutionary model, players can occasionally change their behaviour,
which fits into one of the afore-mentioned three types: conditional cooperation,
unconditional cooperation, or unconditional defection.
The conditional cooperators are further permitted to select a better monitor by voting;
the voters display their preference for a better monitor, from which the
monitors anticipate their potential future payoff if they continue to obey the
present strategy.
We shall see that a cooperative mutualism is achieved if the voters are ready to
select a better monitor in voting rather than change their behaviour in the
donation game.
%
%
%
%


A frequently-studied issue in indirect reciprocity is the evolution of moral
assessment rules which determine what kind of behaviour leads to a good
reputation~\cite{Sigmund2012}.
Well-known assessment rules are SCORING, MILD, and STERN.
The SCORING rule is the simplest assessment rule: cooperation is good and
defection is bad.
Under the MILD and STERN rules, defection against players of bad reputation
(cheaters) is good.
The only disagreement between the MILD and STERN rules is that STERN
prescribes punishing players of bad reputation by withholding help, whereas the
MILD rule leaves both cooperation and defection options open.
The SCORING rule cannot achieve stable cooperation if players simply interact
with one another in random matching games (though the SCORING rule is also known
to stabilise cooperation with some additional assumptions such as players'
growing social networks, multiple reputation states, and assortment in
interactions~\cite{Brandt2005,Tanabe2013,Nax2015}).
The MILD and STERN rules belong to the few that achieve stable cooperation in
random matching games~\cite{Brandt2004,Ohtsuki2004,Ohtsuki2007}.
%
%
%


We study the three above-mentioned assessment rules and find that SCORING
monitors cannot establish cooperative populations, whereas MILD and STERN
monitors can.
When comparing MILD and STERN rules, we find that cooperation has a broader
basin of attraction with the STERN rule.
Moreover, STERN wins when MILD and STERN monitors compete.
However, the MILD rule realises a more cooperative population with less frequent
(and hence, less costly) monitoring than the STERN rule.
This slight difference in the two assessment rules implies a trade-off:
STERN is more stable, but MILD is more efficient.
MILD always wins against SCORING, but SCORING can displace STERN (and thus
subvert cooperation).

\section{Methods}
\label{sec:methods}

Here we summarise the model by which we numerically simulate the
co-evolutionary dynamics.
The derivation of the dynamics is described in more detail in the supporting
information (SI text, Sec.~S1).

\subsection{Population structure, the donation game, and the behaviour of players} 

\begin{figure}[h!]
  \centering
  \includegraphics{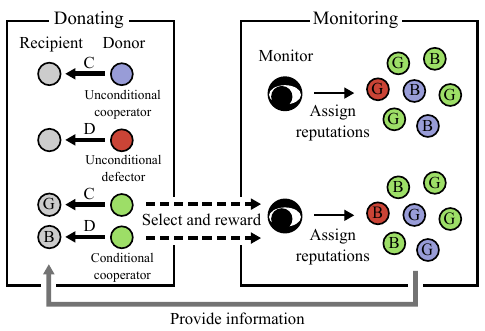}
  \caption{%
    {\bf Schematic overview of the model.}
    We consider donation games among three types of players: unconditional
    cooperators, unconditional defectors, and conditional cooperators.
    Unconditional cooperators always cooperate (C), unconditional defectors
    always defect (D), and conditional cooperators cooperate and defect
    towards recipients with good and bad reputations, respectively.
    The reputation information thus required by conditional cooperators is
    provided to them by monitors in exchange for a fee.
    To allow for competition among different monitoring strategies, we consider
    two monitors who independently observe the players (at a cost to the
    observing monitor) and provide reputation information accordingly (at a cost
    to the requesting conditional cooperator).
    Monitors differ in the fractions of players they observe and in the fees
    they charge for providing information.
    A monitor asked for reputation information about a player who was not
    observed provides a random answer, and each conditional cooperator selects
    either one of the two monitors by comparing the resultant long-term payoffs
    obtained by the monitor's clients.
  }
  \label{fig:schema}
\end{figure}

We consider a large, well-mixed population of players (see
Fig.~\ref{fig:schema}).
From time to time, the players interact with each other in a social dilemma game
called the donation game~\cite{Nowak1998a,Nowak2005}.
In a (one-shot) donation game, two players are selected at random from the
population, and one of them, called the donor, decides whether or not to help
the other, called the recipient.
These two alternatives are called cooperation ($\C$) and defection ($\D$),
respectively.
A donor who cooperates pays a cost $c$ ($> 0$) to increase the recipient's
payoff by an amount $b$ ($> c$).
Each player adopts one of three strategies: unconditional cooperation,
unconditional defection, or conditional cooperation.
An unconditional cooperator or defector always selects $\C$ or $\D$,
respectively.
By contrast, a conditional cooperator selects $\C$ or $\D$ depending on whether
a recipient has a good ($\G$) or bad ($\B$) reputation, respectively.
This reputation information comes at a price $\beta$ ($\ge 0$).

\subsection{Behaviour of monitors} 

A monitor, or information provider, asks a fee, $\beta$, for its service.
It observes each interaction with a probability $q$, for which it has to pay a
cost $C(q) \ge 0$, and updates the record of the player's reputation
accordingly.
%
%
We assume that $C(q)$ is a monotonically increasing convex function such that
the cost is zero with no observation and is infinite with complete observation.
The cost function is proportional to a parameter $\gamma \ge 0$ (see SI text,
Sec.~S1.5).
With probability $1-q$, the monitor records fake information randomly based
on the average ratio of good and bad players in the population.
For example, if 90\% of the players have a good reputation, then a faking
monitor assigns a good reputation to the recipient with a probability of 90\%,
irrespective of the recipient's actual behaviour.
We assume that faking incurs no cost to the monitor.

\subsection{Assessment rules: SCORING, MILD, and STERN} 

A monitor assesses the donor's behaviour according to an assessment rule, which
determines whether the donor obtains a good or a bad reputation (G or B).
We consider three assessment rules called SCORING, MILD, and STERN (see
Tab.~\ref{tab:morals}).
The SCORING rule simply considers that cooperation and defection are good and bad,
respectively, irrespective of the recipient's reputation.
MILD and STERN rules follow the same assessment when the recipient has a good
reputation, whereas they consider that defection against bad players is justified,
\ie, a good behaviour (see $\DB$ column in Tab.~\ref{tab:morals}).
The MILD and STERN rules differ when a donor helps a bad recipient.
Such a behaviour is regarded as good by the MILD rule, whereas it is regarded as
bad by the STERN rule (see $\CB$ column in Tab.~\ref{tab:morals})
We introduce errors in the monitors' assessments.
With a small probability $\mu$, a monitor may assign a reputation opposite to that
intended.
Moreover, we assume that all players have a good reputation to begin with.

\subsection{Social learning among players} 

We study the co-evolution of players and monitors by combining pairwise
comparison and adaptive dynamics, both well established techniques in
evolutionary game theory~\cite{Sandholm2010,Hofbauer1998}.

The players gradually change the relative frequencies of their strategies,
denoted by $(x_\C, x_\D, x_\R)$, where the subscripts denote unconditional
cooperators (C), unconditional defectors (D), and conditional cooperators (R,
for `reciprocators').
Their evolution is driven by an imitation process based on a pairwise payoff
comparison with random exploration, given by
\begin{equation}
  \label{eq:player-dynamics}
  \dot{x}_\sigma = \epsilon \left[\frac{1}{3} - x_\sigma\right] +
        \left(1-\epsilon\right) x_\sigma \sum_{\sigma^\prime} x_{\sigma^\prime}
        \tanh\left[\frac{w}{2} \left(\pi_\sigma - \pi_{\sigma^\prime}\right)\right]
\end{equation}
for each strategy $\sigma \in \{\C, \D, \R\}$, where $\pi_\sigma$ represents the
payoffs of players obeying strategy $\sigma$
(see SI text, Sec.~S1.4 for its derivation).
The first term of the right-hand side of Eq.~\eqref{eq:player-dynamics}
represents random exploration; with a small probability $\epsilon$, the players
explore different strategies in a uniformly random manner.
The second term of the right-hand side of Eq.~\eqref{eq:player-dynamics}
represents imitation based on a pairwise payoff comparison; with a probability
$1-\epsilon$, a randomly selected player compares her payoff and another
randomly selected player's payoff, and imitate the latter player's strategy with
a probability given by a sigmoid function, $1/\left[1 + \exp(-w \Delta)\right]$,
where $\Delta$ is the payoff difference~\cite{Traulsen2006}.
Equation~\eqref{eq:player-dynamics} is tuned by a parameter $w > 0$, which
represents the speed with which players switch to a better strategy~\cite{Traulsen2006}.

\subsection{Voting between monitors and their adaptive dynamics} 

The monitors' evolution is driven by voting by their clients (\ie, conditional
cooperators).
We assume for simplicity that only two monitors, denoted by 1 and 2, are competing.
Most of the time, the two monitors behave alike.
Occasionally, one monitor (monitor 1) slightly changes the parameter values from $(q,
\beta)$ to $(q^\prime, \beta^\prime)$ at random.
The clients of the monitors compare their payoffs, which are different between
the two monitors, and `vote with their hands' on which monitor is better.
That is, the clients show the monitors how many of them will move to a better
monitor, given by
\begin{equation}
  \label{eq:player-softmax-selection}
  \frac{x^\prime_{\R_i}}{x_\R} = \frac{
    \mathrm{e}^{\alpha\pi^\prime_{\R_i}}
  }{
    \mathrm{e}^{\alpha\pi^\prime_{\R_1}} + \mathrm{e}^{\alpha\pi^\prime_{\R_2}}
  }
\end{equation}
for monitor $i \in \{1, 2\}$, if the monitors continue to use the
slightly-changed parameter values (\ie, $(q, \beta)$ and $(q^\prime,
\beta^\prime)$.
Here, $x^\prime_{\R_i} / x_\R$ is the frequency of
clients that vote for monitor $i$ (numerator) relative to the total frequency of
clients (denominator) and $\pi^\prime_{\R_i}$ represents the payoff of clients
that use monitor $i$.
Moreover, the parameter $\alpha > 0$ represents how strongly the clients vote for
the monitor whose clients do better.
This parameter corresponds to how nimbly the monitors evolve their parameters.
On receiving the results of the voting, a less popular monitor, who will lose
some clients in the future if it continues to use the present parameter
values, will quickly follow suit and adopt the more popular monitor's parameter
values.
This process can be modelled by adaptive dynamics (see SI text,
Sec.~S1.5)~\cite{Hofbauer1990}.
The voting is assumed to be much faster than the change in the player's
behaviour from conditional to unconditional cooperation or defection.


\section{Results}
\label{sec:results}


\subsection{The SCORING rule cannot stabilise cooperation} 
\label{sec:results:1st-order-fails}

\begin{figure}[h!]
  \centering
  \includegraphics{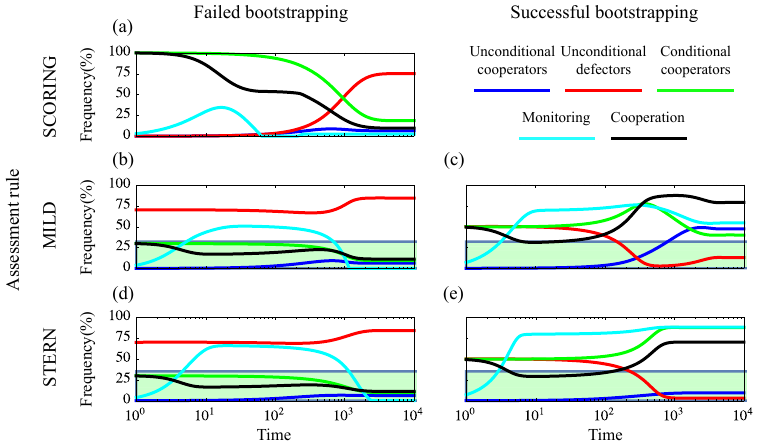}
  \caption{%
    {\bf Failures and successes in the bootstrapping of institutionalised
    monitoring.}
    Bootstrapping occurs when a group without any monitoring gradually evolves
    to exhibit stable and finite levels of monitoring and cooperation.
    Panels show how the frequencies of unconditional cooperators, unconditional
    defectors, and conditional cooperators (blue, red, and green curves,
    respectively), as well as those of monitoring (by monitors; cyan curve) and
    of cooperation (by unconditional or conditional cooperators; black curve)
    evolve from different initial conditions.
    (a) With the SCORING rule, bootstrapping always fails, even for groups
    initially comprised entirely of conditional cooperators.
    (b) With the MILD rule, bootstrapping fails if the initial frequency of
    conditional cooperators is too low (inside the green band).
    (c) With the MILD rule, bootstrapping succeeds if the initial frequency of
    conditional cooperators is high enough (outside of the green band).
    (d) With the STERN rule, bootstrapping fails if the initial frequency of
    conditional cooperators is too low (inside the green band).
    (e) With the STERN rule, bootstrapping succeeds if the initial frequency
    of conditional cooperators is high enough (outside of the green band).
    Within one unit of time, on average, the reputations of all players are
    updated.
    The time axes are scaled logarithmically to show short-term and long-term
    changes together.
    Parameters: $w = 0.01, \alpha = 10, \mu = 0.1, \epsilon = 0.001, \gamma =
    0.01, \kappa = 2, c = 1$, and $b = 10$.
    Initial conditions: $q = 0, \beta = 0, x_\C= 0, x_\D = 1-x_\R$, and $x_\R =
    1$ (a), $x_\R= 0.3$ (b, d), or $x_\R = 0.5$ (c, e).
  }
  \label{fig:examples}
\end{figure}

When both monitors adopt the SCORING rule, the system cannot reach stable
cooperation, even if the initial population of players consists entirely of
conditional cooperators.
Figure~\ref{fig:examples}(a) displays a typical example of the failure of the
SCORING rule.
The frequency of monitoring, \ie, of $x_\R$ and of $q$, first increases.
Then, because the SCORING rule does not distinguish defection against bad players
from defection against good players (\ie, so-called justified defection),
the fraction of good conditional cooperators decreases rapidly, as shown by the
decrease of the frequency of cooperation in Fig.~\ref{fig:examples}(a).
This implies that monitoring harms the population in the case of the SCORING
rule, so that the frequency of monitoring begins to decrease.
Finally, monitoring vanishes and unconditional defectors invade and take over.

\subsection{STERN and MILD rules can stabilise cooperation if voters strongly support a beneficial monitor} 
\label{sec:results:red-king-effect}

When the monitors adopt the MILD or STERN rule, they can secure stable
cooperation supported by frequent monitoring, provided the initial fraction of
conditional cooperators is sufficiently large (Fig.~\ref{fig:examples}(b--e)).
Interestingly, this mutualism between conditional cooperators and monitors is achieved
even if the initial frequency of monitoring is zero, \ie, $q = 0$.
A bootstrapping process allows the monitoring frequency to quickly increase
(see Fig.~\ref{fig:examples}(c,e)).

\begin{figure}[h!]
  \centering
  \includegraphics{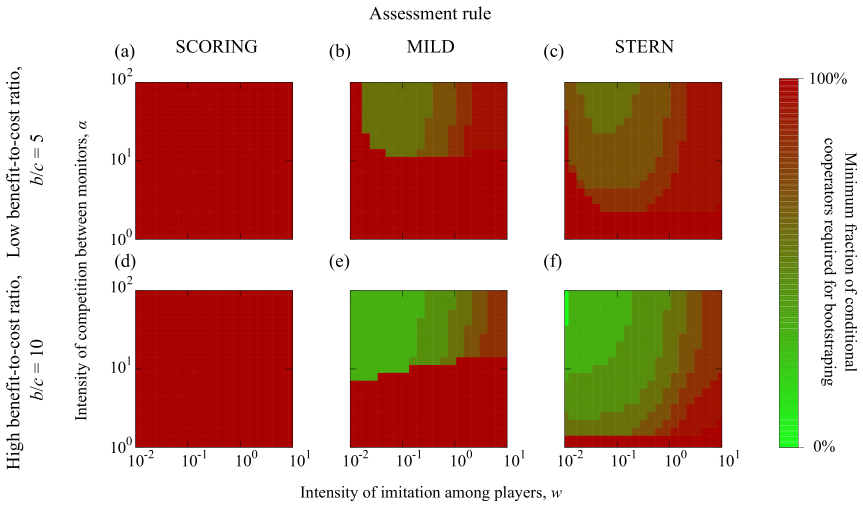}
  \caption{%
    {\bf The bootstrapping of institutionalised monitoring is facilitated by slowly
    evolving players and nimbly adapting monitors.}
    Bootstrapping occurs when a group without any monitoring gradually evolves
    to exhibit stable and finite levels (larger than 10\%) of monitoring and
    cooperation.
    Panels show how the minimum fraction of conditional cooperators required for
    bootstrapping changes with the intensity $w$ of imitation among players and
    the intensity $\alpha$ of competition between monitors.
    Higher intensities imply faster adaptation.
    Low thresholds facilitating bootstrapping are shown in green, and high
    thresholds impeding bootstrapping are shown in red.
    Fully red colouration indicates that bootstrapping is impossible.
    (a,b,c) Low benefit-to-cost ratio of cooperation, $b/c = 5$.
    (d,e,f) High benefit-to-cost ratio of cooperation, $b/c= 10$.
    (a,d) The SCORING rule.
    (b,e) The MILD rule.
    (c, f) The STERN rule.
    Under the SCORING rule, the frequency of monitoring always
    declines to 0, so institutionalised monitoring cannot be established.
    Under the MILD and the STERN rules, bootstrapping is possible and is
    easiest, \ie, requires the least frequency of conditional cooperators, when
    players adapt slowly and monitors adapt quickly.
    Parameters: $\mu= 0.1, \epsilon= 0.001, \gamma = 0.01, \kappa = 2$, and $c =
    1$.
    Initial conditions: $q = 0, \beta = 0, x_\C = 0$, and $x_\D = 1 - x_\R$.
  }
  \label{fig:bootstrap}
\end{figure}

What controls this growth of monitoring is the intensity with which players
select a better monitor in voting (\ie, $\alpha$) relative to that with which
they change their own strategy (\ie, $w$).
We numerically find the minimum fraction of conditional cooperators (\ie, the
minimum $x_\R$) needed to establish a stable mutualism for various values of
$\alpha$ and $w$ (Fig.~\ref{fig:bootstrap}).
In the case of the SCORING rule, as expected, the monitors cannot sustain their
monitoring frequency even if the population consists entirely of conditional
cooperators (Fig.~\ref{fig:bootstrap}(a,d)).
For the MILD and the STERN rules, we find that stable mutualism can be
reached if $\alpha$ is sufficiently large (Fig.~\ref{fig:bootstrap}(b,c,e,f)); a
strong competition between monitors is essential.
Moreover, the required initial fraction of conditional cooperators decreases as
$w$ becomes smaller, provided that the benefit-to-cost ratio of cooperation
(\ie, $b/c$) is sufficiently large (Fig.~\ref{fig:bootstrap}(e,f)).
These two observations together imply that if the voters (\ie, conditional
cooperators) select monitors faster than they switch strategies, then the
monitors are forced to establish reliable monitoring, and thereby the users
enjoy a cooperative society supported by the monitoring system.

\subsection{The STERN rule establishes cooperation more easily than the MILD rule} 
\label{sec:results:stability}

Furthermore, we observe a difference between MILD and STERN; the region
leading to a cooperative mutualism is larger under the STERN rule than under the
MILD rule (compare Fig.~\ref{fig:bootstrap}(b,e) and
Fig.~\ref{fig:bootstrap}(c,f)).
The intensity of competition between monitors (\ie, $\alpha$) required to
reach the cooperative equilibria is larger with the MILD rule than with the
STERN rule.
That is, with a STERN assessment, the system can more easily succeed in
establishing the mutualism, even when the competition between the monitors is
relatively weak.

\subsection{STERN is dominant if STERN and MILD rules compete} 
\label{sec:results:competition}

So far, we have assumed that the two monitors adopt the same assessment rule.
What if different assessment rules compete?
Let us assume that, after a long time over which the two monitors use the same
assessment rule, one of them adopts a different rule, but both monitors still
use the same parameters $q$ and $\beta$.
We can easily see that the payoff to the STERN monitor is always higher than
that to the MILD monitor (see SI text, Sec.~S2).
This is because conditional cooperators using the STERN monitor's information
(STERN users) gain relatively higher payoffs than those using the MILD monitor
(MILD users); when they interact, MILD users cooperate more with STERN users,
whereas STERN users cooperate less with MILD users~\cite{Uchida2010a}.
Thus, STERN is again more robust than MILD, in the sense of the competition
between the two assessment rules~\cite{Pacheco2006,Uchida2010a}.

\subsection{The STERN rule achieves lower cooperation with severe surveillance, whereas the MILD rule achieves higher cooperation with loose monitoring} 
\label{sec:results:efficiency}

\begin{figure}[h!]
  \centering
  \includegraphics{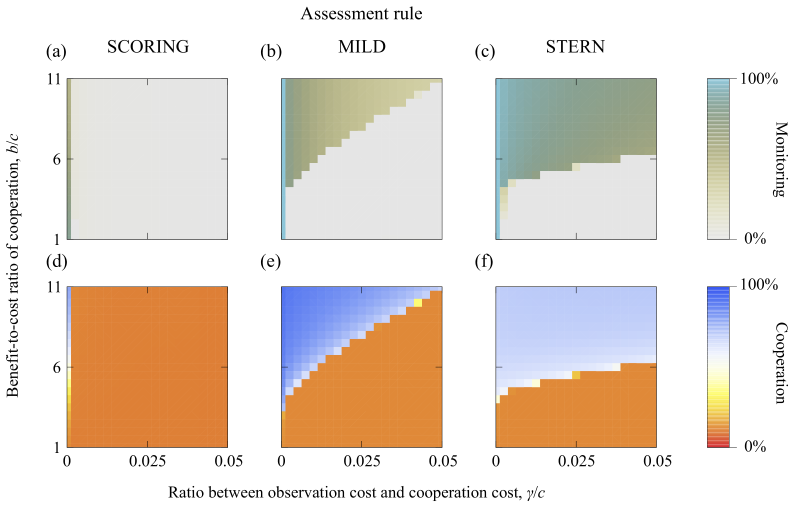}
  \caption{%
    {\bf The MILD rule establishes higher cooperation while requiring only loose
    surveillance, whereas the STERN rule establishes lower cooperation
    while requiring severe surveillance.}
    Panels show how the equilibrium frequencies of (a,b,c) monitoring and
    (d,e,f) cooperation vary with the ratio $\gamma/c$ between observation cost
    and cooperation cost and the benefit-to-cost ratio $b/c$ of cooperation.
    (a,d) The SCORING rule.
    (b,e) The MILD rule.
    (c,f) The STERN rule.
    Under the SCORING rule, the frequency of monitoring always declines to
    $0$, so institutionalised monitoring cannot be established.
    Under the MILD rule, monitor evolution equilibrates at infrequent
    monitoring (loose surveillance) while enabling high frequencies of
    cooperation.
    Under the STERN rule, monitor evolution equilibrates at frequent
    monitoring (severe surveillance) while enabling only intermediate
    frequencies of cooperation.
    In comparison with the MILD rule, the STERN rule is more robust against
    increasing the ratio $\gamma/c$ between observation cost and cooperation
    cost.
    Parameters: $w = 0.01, \alpha = 100, \mu = 0.1, \epsilon = 0.001,
    \kappa = 2$, and $c = 1$.
    Initial conditions: $q = 0, \beta = 0, x_\C = 0, x_\D  = 0$, and $x_\R = 1$.
  }
  \label{fig:equilibria}
\end{figure}

Given a population that has established a stable mutualism, it is interesting to
see whether monitoring is severe or not and how cooperative the players are.
To study this, we numerically observe the equilibrium states of populations
varying in the benefit-to-cost ratio of cooperation in the donation game (\ie,
$b/c$) and in the ratio between monitoring cost and cooperation cost (\ie,
$\gamma/c$) under the two assessment rules MILD and STERN.
The characteristics of equilibria under the three assessment rules differ
qualitatively with respect to the frequency of monitoring
(Fig.~\ref{fig:equilibria}(a,b,c)) and the cooperativeness of the players
(Fig.~\ref{fig:equilibria}(d,e,f)).
In the case of the SCORING rule, again, the monitors cannot increase their monitoring
frequency and the players fail to establish cooperative populations
(Fig.~\ref{fig:equilibria}(a,d)).
In contrast, MILD and STERN rules succeed in establishing cooperative
populations under a wide range of parameter settings (Fig.~\ref{fig:equilibria}(b,c,e,f)).
The equilibrium frequencies of monitoring under MILD and STERN rules are the
same (100\%) when monitoring is cost free (\ie, when $\gamma = 0$; see the left
edges of the panels in Fig.~\ref{fig:equilibria}(a,b)).
When monitoring is costly (\ie, when $\gamma > 0$), one might expect that the
frequency of monitoring would diminish as the cost increases.
This prediction is verified for the MILD rule (Fig.~\ref{fig:equilibria}(a)),
but fails for the STERN rule (Fig.~\ref{fig:equilibria}(b)); in the latter
case, information users still need accurate information although the cost of
monitoring is large.

Why does this happen?
Consider that two STERN monitors have conflicting opinions about a player's
reputation; one monitor (monitor 1) regards the player (player A) as good but
the other monitor (monitor 2) regards the player as bad.
In a donation game, a donor (player B, a conditional cooperator) is informed
about player A's reputation by, say, monitor 1.
Player B helps player A, because player A has a good reputation according to
monitor 1.
In this situation, monitor 1 assigns a good reputation to player B, because the
monitor thinks that the game is in the $\CG$ scenario (see Tab.~\ref{tab:morals}).
However, monitor 2 assigns a bad reputation to player B, because it
thinks that the game is in the $\CB$ scenario.
In this process, the existence of player A, who has conflicting reputations in
the eyes of the two monitors, yields another player who also has conflicting
reputations.
%
Thus the number of players with conflicting reputation inexorably
grows~\cite{Nakamura2012}.
As a consequence, the degree of cooperation under the STERN rule becomes
significantly smaller than that under the MILD rule (Fig.~\ref{fig:equilibria}(e,f)).
To avoid mistakenly cooperating with players that have conflicting reputations,
conditional cooperators need accurate information and require severe
surveillance under the STERN rule.

Another difference between MILD and STERN rules is that in case of the MILD
rule, as the cost of monitoring increases, the minimum benefit-to-cost ratio
(\ie, $b/c$) required for sustaining mutualism becomes larger, whereas in the case
of the STERN rule, it does not change (compare Fig.~\ref{fig:equilibria}(b,e)
with Fig.~\ref{fig:equilibria}(c,f)).
Mutualism under the STERN rule is easier to establish than under the MILD
rule, as previously shown in Fig.~\ref{fig:bootstrap}.

Finally, we mention that if a SCORING monitor competes with a STERN monitor
(both having the same ($q, \beta$)-values), then it may happen that SCORING
wins, thus subverting cooperation (see SI text, Sec.~S3).
This holds if the number of unconditional defectors is sufficiently high.
It follows that under certain conditions, we encounter a rock-paper-scissors
type of competition for the three assessment rules: SCORING beats STERN, MILD
beats SCORING, and STERN beats MILD.

\subsection{Robustness checks} 
\label{sec:results:robustness-checks}

For the results of comparisons between different initial states of players (\ie, $(x_\C,
x_\D, x_\R)$) and different shapes of the cost function for monitoring (\ie,
$C(q)$), see the SI text, Secs.~S3 and S4, respectively.
Neither consideration changes our results qualitatively.
In a few parameter sets unde the MILD rule, we observed stable periodic
oscillations (see the SI text, Sec.~S6 for detail).


\section{Discussion}
\label{sec:discussion}


We have studied a co-evolutionary model of indirect reciprocity in which players
request information about reputations and monitors supply it.
Thus players and monitors mutually benefit from using and providing information.
We compared three different assessment rules called SCORING, MILD and STERN, and
found that only the MILD and STERN rules can establish a cooperative
mutualism.
We confirmed that the SCORING rule fails to foster cooperation
(Sec.~\ref{sec:results:1st-order-fails}).
Mutualism can emerge and be stabilised in the case of the MILD or STERN
rule if the initial frequency of conditional cooperators is sufficiently high
and if they strongly support a better monitor rather than rapidly changing their
strategy; the slow speed of evolution of players' strategy relative to that of
monitors' is important (Sec.~\ref{sec:results:red-king-effect}).
%
The STERN and the MILD rules differ in their stability.
The STERN rule is more robust than the MILD rule in admitting a larger basin of
attraction leading to cooperation (Sec.~\ref{sec:results:stability}).
The intensity of competition between monitors (\ie, $\alpha$) can be smaller in
case of the STERN rule than in the case of the MILD rule.
The STERN rule is more robust than the MILD rule in another sense: the
competition between two monitors, one STERN and one MILD, always leads to
victory by the STERN rule (Sec.~\ref{sec:results:competition}).
Moreover, the difference between the MILD and the STERN rules substantially
affects the outcome of co-evolution.
With MILD monitors, players achieve more cooperative states under less-frequent
monitoring, whereas with STERN monitors, players achieve less cooperative states
and are under severe surveillance, \ie, $q \approx 1$
(Sec.~\ref{sec:results:efficiency}).
However, cooperative mutualism can be more easily obtained with STERN
monitors than with MILD monitors in the sense that the cost-to-benefit ratio and
the cost for monitoring can be larger.


In evolutionary studies of symbiosis, the so-called Red Queen's hypothesis is
often invoked.
It says that competing species are exposed to arms races and therefore those
evolving faster are advantaged~\cite{VanValen1973,Dawkins1979}.
However, recent theoretical studies have found that sometimes the species
evolving slowly can win.
This is called the Red King effect~\cite{Bergstrom2003,Damore2011}.
In the Red King effect, immobility can be a form of commitment that obliges
other species to give way.
In the present study, a similar effect enables a stable mutualism between
players and monitors; players are the hosts that evolve slowly and promote
the monitors' costly monitoring.


%
%
%
%


Several works in economics have studied repeated games with costly monitoring of
opponents'
actions~\cite{Ben-Porath2003,Miyagawa2008,Flesch2009,Fujiwara-Greve2012}.
These studies focused on the individual trade-off between the value of
information and the cost of its acquisition and did not consider how to
promote costly sharing of information among individuals.
Gazzale presented a model of seller--buyer transaction in which buyers can
report information about sellers to a rating system and their reporting is
visible by sellers, and Gazzale and Khopkar experimentally studied how this
mechanism promotes costly sharing of information~\cite{Gazzale2005,Gazzale2011}.
In their model, a buyer's costly reporting of information about a seller builds
the buyer's reputation as an information spreader.
This increases the effort level of the buyer's future partners afraid of receiving
a bad reputation, and thus buyers have an incentive to report information even if
it is costly to do so.
In our model, instead, monitors make an effort because by doing so their
information users reward them.


The above-mentioned studies did not assume that the reported information may be
fake and that deceivers who shirk costly monitoring gain more than serious
information providers.
This problem of spreading false information about reputations was, as far as
we know, first studied in biology by Nakamaru and Kawata~\cite{Nakamaru2004}.
%
%
In their study, a `conditional advisor' was capable of detecting and suppressing
free-riding liars.
This is a strategy by which a player (player A) spreads reputation information
about others, which is received from another player (player B) only when B had
previously cooperated with A.
The conditional advisor strategy, therefore, needs a large amount of information
acquisition for the verification of reputation information.
In contrast, our model does not require individuals to verify their information;
they only need to select a more beneficial monitor.
This implies that information users can trust information providers more easily
when the providers are exposed to competition with each other.
%


Rockenbach and Sadrieh conducted a behavioural experiment on the subject of
costly information spreading~\cite{Rockenbach2012}.
They demonstrated that people tend to share helpful information with others even
if reporting it provides no individual benefit.
Such an instinct for the acquisition and sharing of information could evolve if
it is usually rewarded~\cite{Rand2013}.
In our model, we assumed that all individuals including players and monitors
are only motivated by self-interest.
We demonstrated theoretically that the reward for reporting helpful information
can overcome the problem of costly information acquisition, even if individuals
have no social preferences other than pure self-interest.
%


The present study is restricted to a simple model, and the following extensions
would provide further insights.
First, we studied competition between two monitors only, rather than between many.
In real life, situations with more than two competitors are common, and `hub'
individuals with huge numbers of connections on social networks are
observed~\cite{Newman2003}.
Whether a hub information provider emerges from competition among many monitors
or not is an interesting question.
%
%
%
Second, we assumed that when monitors fail to engage in costly observation, they
deceive client players by faking random information.
In real life, such falsification might be strategic;
%
%
for example, monitors might be corrupted by players offering them money for
reporting a good reputation~\cite{Masclet2012}.
Third, we showed that the competition between monitors driven by their clients'
voting `by hands' rather than `by feet' enables cooperation;
clients only show their preference over monitors under voting by hands,
whereas they actually move to a better monitor under voting by feet.
This is in contrast to most studies of evolutionary dynamics, which typically
assume voting by feet.
If monitors compete under voting by feet, it seems likely that one monitor could
take the entire of the clients, even if they used the same parameters.
Therefore, it is important to study whether cooperation emerges if clients vote
with their feet as well as the difference between the two types of voting.
Fourth, our model assumed that social learning among players occurs in a
well-mixed manner, \ie, that the population does not have structure.
However, it could be the case that a population has a structure; people may
learn from their neighbours~\cite{Perc2009}.
In that case, cooperation might be established even if the initial fraction of
conditional cooperators is smaller than that in the present result (see
Fig.~\ref{fig:bootstrap}).
This is because a structure increases clustering of players having the
same strategy and helps cooperation\cite{Nowak2010}.
Fifth, in our model, we only introduced errors in the monitors' assessments,
which yielded conflicting opinions about a player's reputation and thus players
under the STERN rule were less cooperative than those under the MILD rule.
To introduce other types of errors, \eg, errors in each player's perception
about reputation-related information, increases such conflicting opinions and
therefore it could reduce cooperation more.


An important characteristic of human behaviour is the ability to establish
large-scale cooperation~\cite{Fehr2004}.
Such large-scale cooperation partially depends upon the development of
large-scale information sharing, which suffers from a tragedy of the commons.
As we have discussed, one possibility for overcoming this dilemma is to introduce
competition between information sharing systems.
We hope that this study helps to build understanding of sustainable mechanisms
for information provision under indirect reciprocity.

\section*{Authors' contributions}

MN carried out the mathematical analysis.
MN and UD conceived of the study, designed the study, and wrote the
manuscript.
All authors gave final approval for publication.

\section*{Competing interests}

We have no competing interests.

\section*{Funding}

MN gratefully acknowledges support by JSPS KAKENHI Grant No.~13J05595.
UD gratefully acknowledges support by the Austrian Science Fund (FWF), through
a grant for the research project {\it The Adaptive Evolution of Mutualistic
Interactions} (TECT I-106 G11) as part of the multi-national collaborative
research project {\it Mutualisms, Contracts, Space, and Dispersal}
(BIOCONTRACT) selected by the European Science Foundation as part of the
European Collaborative Research (EUROCORES) Programme {\it The Evolution of
Cooperation and Trading} (TECT).
UD gratefully acknowledges additional support by the European Science
Foundation (EUROCORES), and the European Science Foundation (ESF).

\section*{Acknowledgments}

We thank Karl Sigmund for valuable discussions throughout this work.

\bibliographystyle{prsb}
\bibliography{refs}

\clearpage
\section*{Figures}

\clearpage

\clearpage
\section*{Tables}

\begin{table}[h!]
  \caption{%
    {\bf Assessment rules.}
    When observing a donation game, each monitor assigns a reputation, either
    good (G) or bad (B), to the participating donor according to an assessment
    rule (SCORING, MILD, or STERN).
    These assessment rules differ in the four social scenarios:
    $\CG$, $\DG$, $\DB$, and $\CB$.
    In the $\CG$ scenario, a donor cooperates with a good recipient,
    in the $\DG$ scenario, a donor defects against a good recipient,
    in the $\DB$ scenario, a donor defects against a bad recipient,
    and in the $\CB$ scenario, a donor cooperates with a bad
    recipient.
    In the table, each cell represents the reputation that the donor receives
    in each scenario under the two assessment rules.
    The SCORING rule regards cooperating ($\CG$ and $\CB$) donors as good and
    defecting ($\DG$ and $\DB$) donors as bad.
    The MILD and the STERN rules are the same except for the cell $\CB$;
    they regard the donor in this scenario as good and bad, respectively.
  }
  \label{tab:morals}
  \centering
  \begin{tabular}{|c|cccc|}
    \hline
    \multirow{2}{*}{\makebox[30mm][c]{Assessment rule}} &
    \multicolumn{4}{c|}{Social scenario} \\
    & \makebox[20mm][c]{$\CG$} &
      \makebox[20mm][c]{$\DG$} &
      \makebox[20mm][c]{$\DB$} &
      \makebox[20mm][c]{$\CB$} \\
    \hline
    SCORING & Good & Bad & Bad  & Good \\
    MILD    & Good & Bad & Good & Good \\
    STERN   & Good & Bad & Good & Bad  \\
    \hline
  \end{tabular}
\end{table}



\section*{Supplementary material}

# Supplementary Information: Voting by Hands Promotes Institutionalised Monitoring in Indirect Reciprocity

Mitsuhiro Nakamura    Ulf Dieckmann

October 6, 2016

## S1 Model description

This section complements the model description in detail. In the model, monitors continuously assign reputations to players. We first introduce the dynamics of the reputation assignments (Sec. S1.1). Based on the distribution of reputations of players, their expected payoff (Sec. S1.2) and the frequency of cooperation (Sec. S1.3) are calculated. We next introduce the social learning dynamics of the players (Sec. S1.4) as well as that of monitors (Sec. S1.5). Throughout, we assume that the learning of monitors is sufficiently faster than that of players. This allows us to consider that in order to derive the dynamics of players, the two monitors adopt a unique strategy $s = (q, \beta)$ (Secs. S1.1, S1.2, S1.4). To derive the dynamics of the monitors, we assume that they slightly differentiate their strategies for a moment (Sec. S1.5). Table S1 summarises the definitions of symbols used in this section.

### S1.1 Reputation dynamics

We analyse the dynamics of the reputation distribution, *i.e.*, the fractions of good and bad players in the eyes of the two monitors. From time to time, two players are selected from a population at random, and they play a donation game. From the game, the two monitors, denoted by 1 and 2, assign reputations to the donor based on either actual observation or fake information produced in a random manner. We consider the dynamics of the distribution of players that have reputation vectors $r = (r_1, r_2) \in \{G, B\}^2$ in the eyes of the two monitors. Extending the method proposed by Ref. [1], the dynamics are represented by the following equation for each $r$:

$$\dot{p}(r) = -p(r) + \sum_{\sigma \in \{C, D, R_1, R_2\}} x_\sigma p_\sigma(r), \tag{S1}$$

where the dot denotes the time derivative, $R_i$ for $i \in \{1, 2\}$ represents conditional cooperators that use the reputation information provided by each monitor $i$, $x_\sigma$ represents the fraction of players that

1

Table S1: **Meaning of symbols.**

| Symbol | Meaning |
|---|---|
| $x_C$ | Fraction of unconditional cooperators |
| $x_D$ | Fraction of unconditional defectors |
| $x_R$ | Fraction of conditional cooperators |
| $x_{R_1}$ | Fraction of conditional cooperators using the monitor 1 |
| $x_{R_2}$ | Fraction of conditional cooperators using the monitor 2 |
| $q$ | Frequency of monitoring |
| $\beta$ | Information fee required by the monitors |
| $c$ | Cost of cooperation |
| $b$ | Benefit of cooperation |
| $w$ | Intensity of imitation between players |
| $\alpha$ | Intensity of monitor-selection by unconditional cooperators |
| $\mu$ | Probability that a monitor mistakenly assigns a reputation |
| $\epsilon$ | Probability that a player randomly changes his/her strategy regardless of pairwise payoff comparison |
| $r$ | Player's reputation vector, GG, GB, BG, or BB, in the eyes of the two monitors |
| $p(r)$ | Fraction of players having a reputation vector $r$ |
| $p_\sigma(r)$ | Fraction of players adopting a strategy $\sigma$ and having a reputation vector $r$ |
| $a_\sigma(r)$ | Action (C or D) of a donor, who adopts strategy $\sigma$, when playing a game with a recipient having a reputation vector $r$ |
| $\delta_G(a, r)$ | Probability that a monitor assigns a good reputation to a donor selecting an action $a$ toward a recipient having a reputation $r$ (with monitoring) |
| $\rho$ | Probability that a monitor randomly assigns a good reputation according to the average ratio of good and bad players (without monitoring) |

adopt a strategy $\sigma$, and $p_\sigma(r)$ represents the probability that a player adopting strategy $\sigma$ receives a reputation vector, $r$, in a one-shot donation game. $p_\sigma(r)$ is given by

$$p_\sigma(r) = \sum_{r' \in \{G,B\}^2} p(r')\phi_{r_1}^s(a_\sigma(r'), r'_1)\phi_{r_2}^s(a_\sigma(r'), r'_2), \qquad (S2)$$

where a donor's action, $a_\sigma(r')$, depends upon his/her strategy $\sigma$ such that $a_C(r') = C$ for any $r'$ (unconditional cooperation), $a_D(r') = D$ for any $r'$ (unconditional defection), $a_{R_1}(r') = \eta(r'_1)$ (conditional cooperation based on the information provided by monitor 1) and $a_{R_2}(r') = \eta(r'_2)$ (conditional cooperation based on the information provided by monitor 2) with $\eta(G) = C$ and $\eta(B) = D$. $\phi_r^s(a, r')$ represents the probability that a monitor adopting a strategy $s$ assigns a reputation $r$ to a donor that has selected action $a$ towards a recipient having a reputation $r'$ in the eyes of the monitor. $\phi_r^s(a, r')$ is given by

$$\phi_G^s(a, r) = q\delta_G(a, r) + (1 - q)\rho \qquad (S3)$$

and $\phi_B^s(a, r) = 1 - \phi_G^s(a, r)$, where $q$ is the probability that a monitor does monitoring, and $\delta_G(a, r)$ and $\rho$ represent the probabilities with which a monitor assigns a good reputation to a donor after

2

doing and skipping monitoring, respectively. In Tab. S2, we list the $\delta_G$ values in Eq. (S3) under each assessment rule. Hereafter, we use the following notations for the reputation vectors: $GG \equiv (G, G)$,

Table S2: **Probabilities that a monitor assigns a good reputation to a donor after monitoring.** $\mu$ represents the probability that the reputation is assigned erroneously because of an assessment error.

| Assessment rule | $\delta_G(C, G)$ | $\delta_G(D, G)$ | $\delta_G(C, B)$ | $\delta_G(D, B)$ |
|---|---|---|---|---|
| SCORING | $1 - \mu$ | $\mu$ | $1 - \mu$ | $\mu$ |
| MILD | $1 - \mu$ | $\mu$ | $1 - \mu$ | $1 - \mu$ |
| STERN | $1 - \mu$ | $\mu$ | $\mu$ | $1 - \mu$ |

$GB \equiv (G, B)$, $BG \equiv (B, G)$, and $BB \equiv (B, B)$. We assume that the monitors adjust their $\rho$ values to the average fraction of good players in the population, *i.e.*, $\rho = p(G*) \equiv p(GG) + p(GB)$ in the eyes of monitor 1 and $\rho = p(*G) \equiv p(GG) + p(BG)$ in the eyes of monitor 2. Because here the monitors adopt the same strategy and the same assessment rule, the fractions of $R_1$- and $R_2$-players are equal, *i.e.*, $\rho = p(G*) = p(*G)$ holds true.

## S1.2 Payoff of players when monitors adopt the same strategy

We denote the expected payoff of a player that adopts a strategy $\sigma$ by

$$\pi_\sigma = -\beta_\sigma - ch_\sigma + bg_\sigma, \tag{S4}$$

where $\beta_\sigma$ represents the information fee, $h_\sigma$ represents the probability that the player helps (*i.e.*, selects C towards) a random recipient, and $g_\sigma$ represents the probability that the player is helped (*i.e.*, receives C) by a random donor. Clearly, $\beta_C = \beta_D = 0$, $\beta_{R_1} = \beta_{R_2} = \beta$, $h_C = 1$, and $h_D = 0$. $h_{R_1}$ and $h_{R_2}$ depend upon the fractions of good players in the eyes of the two monitors, which are given by

$$h_{R_1} = p(G*) \tag{S5a}$$

and

$$h_{R_2} = p(*G). \tag{S5b}$$

Note that, since the two monitors adopt the same strategy and the same assessment rule, $p(G*) = p(*G) \iff h_{R_1} = h_{R_2}$ holds true. $g_\sigma$ is given by

$$g_\sigma = x_C + x_{R_1} p_\sigma(G*) + x_{R_2} p_\sigma(*G). \tag{S6}$$

As above, $x_{R_1} = x_{R_2} = x_R/2$ and $\pi_{R_1} = \pi_{R_2}$ hold true. Thus, $\pi_R = (x_{R_1}\pi_{R_1} + x_{R_2}\pi_{R_2})/x_R = \pi_{R_1} = \pi_{R_2}$.

## S1.3 Frequency of cooperation among players when monitors adopt the same strategy

The frequency of cooperation among the players is given by

$$x_C + x_R p(G*), \tag{S7}$$

where cooperators always cooperate (the first term in Eq. (S7)) and conditional cooperators cooperate with players that have good reputations in the eyes of either of the two monitors (the second term in Eq. (S7)). Note that $p(G*) = p(*G)$ because here the two monitors adopt the same strategy.

## S1.4 Social learning dynamics of players

For the strategy updates of players, we employ pairwise comparison dynamics that describe the so-cial learning process among players based on their payoff comparison and random exploration [2]. From time to time, a randomly selected player has a chance to change his/her strategy. With a probability $\epsilon$, the player changes his/her strategy uniformly at random. With a probability $1 - \epsilon$, the player refers to another randomly selected player's payoff and imitates the referred player's strategy with a probability $s(w\Delta)$, where $s(x) = 1/[1 + \exp(-x)]$ is a sigmoid function, $\Delta$ is the payoff difference between the two players, and parameter $w \geq 0$ controls the intensity of imitation when a player tries to imitate another player's strategy. We assume that within a unit time interval, all players in the population update their strategies. With a time interval $\Delta t$, the fraction of players obeying strategy $\sigma$ changes on average to

$$x_\sigma(t + \Delta t) = (1 - \Delta t)x_\sigma(t) + \epsilon \Delta t \sum_{\sigma'} W^{\text{random}}_{\sigma' \to \sigma} + (1 - \epsilon)\Delta t \sum_{\sigma'} W^{\text{imitation}}_{\sigma' \to \sigma}, \tag{S8}$$

where $W^{\text{random}}_{\sigma' \to \sigma} = 1/3$ represents the probability with which a player changes his/her strategy from $\sigma'$ to $\sigma$ by a random pick, and $W^{\text{imitation}}_{\sigma' \to \sigma}$ represents the probability with which a player changes his/her strategy from $\sigma'$ to $\sigma$ by a payoff comparison. $W^{\text{imitation}}_{\sigma' \to \sigma}$ is given by

$$W^{\text{imitation}}_{\sigma' \to \sigma} = \begin{cases} x_\sigma + \sum_{\sigma'' \neq \sigma} x_{\sigma''} s(-w(\pi_{\sigma''} - \pi_\sigma)), & \text{if } \sigma' = \sigma \quad \text{(S9a)} \\ x_\sigma s(w(\pi_\sigma - \pi_{\sigma'})), & \text{if } \sigma' \neq \sigma \quad \text{(S9b)} \end{cases}$$

where $\pi_\sigma$ and $\pi_{\sigma'}$ are the players' payoff (Eq. (S4)). In Eq. (S9a), a focal player obeying strategy $\sigma$ stays at the same strategy $\sigma$ when the player refers to a player obeying the same strategy $\sigma$ (with a probability $x_\sigma$) or when the player refers to a player obeying another strategy $\sigma''$ (with probability $x_{\sigma''}$) and does not imitate the strategy $\sigma''$ (with a probability $s(-w(\pi_{\sigma''} - \pi_\sigma))$). In Eq. (S9b), a focal player obeying strategy $\sigma'$ other than $\sigma$ imitates the strategy $\sigma$ when the focal player refers to a player obeying strategy $\sigma$ (with a probability $x_\sigma$) and imitates it (with a probability $s(w(\pi_\sigma - \pi_{\sigma'}))$). Taking the limit $\Delta t \to 0$ in Eq. (S8), we obtain

$$\dot{x}_\sigma = \epsilon \left[ \frac{1}{3} - x_\sigma \right] + (1 - \epsilon) x_\sigma \sum_{\sigma'} x_{\sigma'} \tanh\left[ \frac{w}{2} (\pi_\sigma - \pi_{\sigma'}) \right] \tag{S10}$$

for each strategy $\sigma \in \{C, D, R\}$. Note that here we consider $x_R = x_{R_1} + x_{R_2}$.

4

## S1.5 Social learning dynamics of monitors

Now let us consider that for a moment, monitor 1 slightly changes his/her strategy from $s = (q, \beta)$ to $s' = (q', \beta')$. In this situation, the $R_1$- and $R_2$-players obtain different payoffs. We assume that, temporarily, the payoffs of $R_1$- and $R_2$-players are changed to

$$\pi'_{R_1} = -\beta' - ch'_{R_1} + bg'_{R_1} \tag{S11a}$$

and

$$\pi'_{R_2} = -\beta - ch'_{R_2} + bg'_{R_2}, \tag{S11b}$$

respectively, where

$$h'_{R_1} = p'(G*), \tag{S12a}$$
$$h'_{R_2} = p'(*G), \tag{S12b}$$

and

$$g'_\sigma = x_C + x_{R_1} p'_\sigma(G*) + x_{R_2} p'_\sigma(*G) \tag{S12c}$$

for $\sigma \in \{R_1, R_2\}$. In Eq. (S12), $p'(r)$ and $p'_\sigma(r)$ for $r \in \{GG, GB, BG, BB\}$ represent the transient reputation distribution while the two monitors differentiate their strategies, which are given by

$$p'(r) = \sum_\sigma x_\sigma p'_\sigma(r) \tag{S13a}$$

and

$$p'_\sigma(r) = \sum_{r' \in \{G,B\}^2} p(r') \phi^{s'}_{r_1}(a_\sigma(r'), r'_1) \phi^s_{r_2}(a_\sigma(r'), r'_2). \tag{S13b}$$

Based on the transient payoffs, i.e., Eq. (S11), all $R_1$- and $R_2$-players simultaneously vote for the monitors. The fractions of votes by the $R_1$- and $R_2$-players are given by a softmax function,

$$\frac{x'_{R_i}}{x_R} = \frac{e^{\alpha \pi'_{R_i}}}{e^{\alpha \pi'_{R_1}} + e^{\alpha \pi'_{R_2}}} \tag{S14}$$

for $i \in \{1, 2\}$, where $\alpha > 0$ controls the intensity of the information users' preference to vote for a better monitor. Note that $\sum_i x'_{R_i} = x_R$ holds true.

After the voting, the monitors consider that, if they continue to stay in their strategies $s'$ resp. $s$, their clients will actually change their shares to $x'_{R_1}$ resp. $x'_{R_2}$. The monitors expect that their payoffs will change to

$$P(s') = -C(q') + \beta' x'_{R_1} \tag{S15a}$$

5

for monitor 1 and

$$P(s) = -C(q) + \beta x'_{R_2}, \tag{S15b}$$

for monitor 2, where we assume the observation cost function as

$$C(q) = \gamma' \left[ (1-q)^{-\kappa'} - 1 \right] \tag{S16}$$

with $\gamma' = \gamma/(\kappa - 1)$, $\kappa' = \kappa - 1$, $\gamma \geq 0$, and $\kappa > 1$. Clearly, $C(0) = 0$ and $C(1) = \infty$ when $\gamma > 0$. We define the cost function as it is in order to simplify its gradient:

$$\frac{dC(q)}{dq} = \gamma(1-q)^{-\kappa}. \tag{S17}$$

We consider that, before the information users actually change their shares, one of the two monitors stochastically imitates the other in the manner of a pairwise comparison, which depends on the difference between their payoffs, given by

$$P(s') - P(s) = C(q) - C(q') + x_R \frac{\beta' e^{\alpha \pi'_{R_1}} - \beta e^{\alpha \pi'_{R_2}}}{e^{\alpha \pi'_{R_1}} + e^{\alpha \pi'_{R_2}}}. \tag{S18}$$

We approximate the evolution of the monitors' strategy by

$$\frac{d\tau}{dt} = \left. \frac{\partial [P(s') - P(s)]}{\partial \tau'} \right|_{s'=s}, \tag{S19}$$

where $(\tau, \tau')$ is either $(q, q')$ or $(\beta, \beta')$. This yields

$$\begin{cases} \dot{q} = -\gamma(1-q)^{-\kappa} + \alpha\beta \dfrac{x_R}{2} \left. \dfrac{\partial(\pi'_{R_1} - \pi'_{R_2})}{\partial q'} \right|_{s'=s}, \\[2ex] \dot{\beta} = \dfrac{x_R}{2} + \alpha\beta \dfrac{x_R}{2} \left. \dfrac{\partial(\pi'_{R_1} - \pi'_{R_2})}{\partial \beta'} \right|_{s'=s}. \end{cases} \tag{S20}$$

We calculate Eq. (S20) under each assessment rule. Note that for the gradient of $q$ in numerical simulations, because $q$ is bounded between 0 and 1, we use

$$\begin{cases} \max(0, \dot{q}), & \text{for } q \leq h, \\ \dot{q}, & \text{for } h < q \leq 1 - h, \\ \min(0, \dot{q}), & \text{for } 1 - h < q, \end{cases} \tag{S21}$$

where $h = 0.001$ is a fixed threshold.

Because $\partial(\pi'_{R_1} - \pi'_{R_2})/\partial\beta'\big|_{s'=s} = -1$ in Eq. (S20), the gradient of $\beta$ is simply reduced to

$$\dot{\beta} = \frac{x_R}{2} (1 - \alpha\beta). \tag{S22}$$

6

Therefore, $\beta$ always converges to a unique value $\beta^* \equiv 1/\alpha$ if $x_R > 0$.

The gradient of $q$ depends upon the assessment rule and is non-trivial. Here we want to determine the analytical form of $\partial(\pi'_{R_1} - \pi'_{R_2})/\partial q'|_{s'=s}$ in Eq. (S20). From Eq. (S11), we see that

$$\frac{\partial(\pi'_{R_1} - \pi'_{R_2})}{\partial q'}\bigg|_{s'=s} = -c\,\frac{\partial(h'_{R_1} - h'_{R_2})}{\partial q'}\bigg|_{s'=s} + b\,\frac{\partial(g'_{R_1} - g'_{R_2})}{\partial q'}\bigg|_{s'=s}, \tag{S23}$$

where

$$\frac{\partial(h'_{R_1} - h'_{R_2})}{\partial q'}\bigg|_{s'=s} = \frac{\partial}{\partial q'}\left[p'(GB) - p'(BG)\right]\bigg|_{s'=s}$$

$$= \frac{\partial}{\partial q'}\sum_\sigma x_\sigma \sum_{r'} p(r')\left[\phi_G^{s'}(a_\sigma(r'), r'_1)\phi_B^s(a_\sigma(r'), r'_2) - \phi_B^{s'}(a_\sigma(r'), r'_1)\phi_G^s(a_\sigma(r'), r'_2)\right]\bigg|_{s'=s}$$

$$= \sum_\sigma x_\sigma \sum_{r'} p(r')\delta_G(a_\sigma(r'), r'_1) - \rho \tag{S24a}$$

and

$$\frac{\partial(g'_{R_1} - g'_{R_2})}{\partial q'}\bigg|_{s'=s} = \frac{\partial}{\partial q'}\frac{x_R}{2}\left[(p'_{R_1}(G*) + p'_{R_1}(*G)) - (p'_{R_2}(G*) + p'_{R_2}(*G))\right]\bigg|_{s'=s}$$

$$= \frac{x_R}{2}\frac{\partial}{\partial q'}\left[p'_{R_1}(G*) - p'_{R_2}(G*)\right]\bigg|_{s'=s}$$

$$= \frac{x_R}{2}\sum_{r'} p(r')\left[\delta_G(\eta(r'_1), r'_1) - \delta_G(\eta(r'_2), r'_1)\right]. \tag{S24b}$$

Applying Eq. (S23) to Eq. (S20) and using Tab. S2, we obtain

$$\dot{q} = -\gamma(1-q)^{-\kappa} + \alpha\beta\frac{x_R}{2}\left(-cA + bB\right), \tag{S25}$$

where $A$ and $B$ are

$$\begin{cases} A &= x_C(1-\mu) + x_D\mu + \\ & x_R[\mu + (1-2\mu)p(G*)] - p(G*), \\ B &= 0, \end{cases} \tag{S26}$$

in the case of SCORING,

$$\begin{cases} A &= x_C(1-\mu) + x_D\left[1 - \mu - p(G*)(1-2\mu)\right] + \\ & x_R[1 - \mu - \frac{1}{2}(1-2\mu)p(GB)] - p(G*), \\ B &= \frac{1}{2}x_R(1-2\mu)p(GB), \end{cases} \tag{S27}$$

in the case of MILD, and

$$\begin{cases} A &= x_C\left[\mu + p(G*)(1-2\mu)\right] + x_D\left[1 - \mu - p(G*)(1-2\mu)\right] + \\ & x_R\left[1 - \mu - (1-2\mu)p(GB)\right] - p(G*), \\ B &= x_R(1-2\mu)p(GB), \end{cases} \tag{S28}$$

in the case of STERN.

7

## S2 STERN dominates MILD when two monitors adopting the two assessment rules compete against each other

Here, we analyse a competition between two monitors that have the same parameters (*i.e.*, $q$ and $\beta$) but different assessment rules, either MILD or STERN. We consider that, after a long time in which two monitors adopt an identical assessment rule, one of them changes its rule to the other one, *i.e.*, STERN (MILD) if the two monitors have adopted MILD (STERN). We denote that the monitors 1 and 2 adopt STERN and MILD, respectively. In this situation, the difference in payoffs between the two monitors is given by

$$P(s_1|s_2) - P(s_2|s_1) = \beta x_R \tanh\left[\frac{\alpha}{2}\left(\pi'_{R_1} - \pi'_{R_2}\right)\right], \tag{S29}$$

where $\pi'_{R_1} - \pi'_{R_1}$ is given by Eq. (S11) with the assumption that the monitors 1 and 2 use the STERN and MILD rules, respectively (see Tab. S2). If $\beta x_R$ is positive, the sign of Eq. (S29) is the same as that of $\pi'_{R_1} - \pi'_{R_1}$. A straightforward calculation leads to

$$\pi'_{R_1} - \pi'_{R_2} = q(1 - 2\mu)\left[\frac{x_R}{2}p(\text{BG})(b + c) + x_C p(\text{B}*)c\right]. \tag{S30}$$

Note that to derive Eq. (S30), we use $p(\text{GB}) = p(\text{BG})$ because the distribution of reputations is symmetric before one of the two monitors changes its assessment rule. Equation (S30) is positive if $q > 0$. Therefore, if there are at least a small fraction of conditional cooperators and the probability of monitoring is not zero, the payoff of the STERN monitor is better than that of the MILD monitor.

Equation (S30) is proportional to a weighted summation of $x_R/2 \cdot p(\text{BG})$ and $x_C p(\text{B}*)$. Intuitively, the former term is yielded when conditional cooperators using the MILD monitor cooperate with recipients with a bad reputation in the eyes of the STERN monitor. The latter term is yielded when unconditional cooperators cooperate with recipients with a bad reputation in the eyes of the STERN monitor. That is, conditional cooperators using the MILD monitor or unconditional cooperators are punished by the STERN monitor since they help ill-reputed players in the eyes of the STERN monitor [3].

## S3 SCORING can beat STERN if the frequency of unconditional defectors is sufficiently high

Here, we consider a competition between SCORING and STERN monitors that have the same parameters (*i.e.*, $q$ and $\beta$). The situation is the same as in Sec. S2, except that the monitors 1 and 2 adopt SCORING and STERN, respectively. In this case, a straightforward calculation leads to

$$\pi'_{R_1} - \pi'_{R_2} = q(1 - 2\mu)\left[-x_R p(\text{GB})(b + c) + (1 - 2x_C)p(\text{B}*)c\right]. \tag{S31}$$

Equation (S31) is positive if $x_C < 1/2$ and

$$\frac{b}{c} < \frac{1 - 2x_C}{x_R}\frac{p(\text{B}*)}{p(\text{GB})} - 1. \tag{S32}$$

8

Thus, given a fixed benefit-to-cost ratio of cooperation, *i.e.*, $b/c$, if there are a sufficiently small fraction of unconditional and conditional cooperators, *i.e.*, if there are a sufficiently large fraction of unconditional defectors, Eq. (S32) is satisfied and a monitor has an incentive to adopt SCORING over STERN.

## S4 Effects of the initial state of players upon the outcome

In the main text, we assumed that the initial state of players is somewhere on the edge between a monomorphism of conditional cooperators and that of defectors, *i.e.*, $(x_C, x_D, x_R) = (0, 1 - x, x)$, where $x \in [0, 1]$ varies. To complement the main results, in Fig. S1, we show the outcomes of

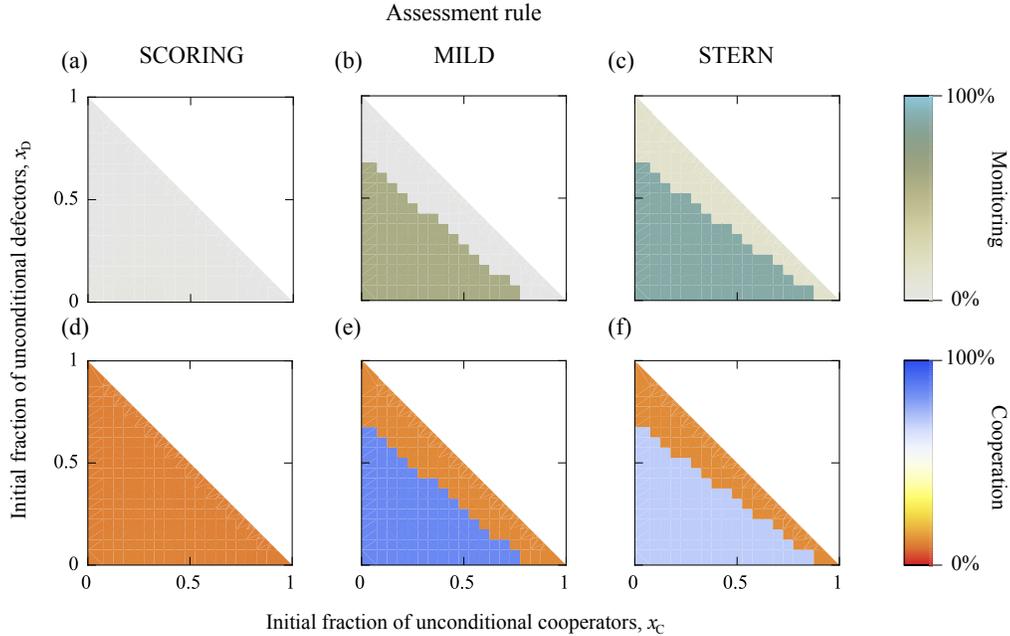

Figure S1: **Outcomes of co-evolution when starting from various initial states of the population of players.** Panels show the equilibrium frequencies of **(a,b,c)** monitoring and **(d,e,f)** cooperation when varying the initial fractions of unconditional cooperators ($x_C$) and unconditional defectors ($x_D$). **(a,d)** The SCORING rule. **(b,e)** The MILD rule. **(c,f)** The STERN rule. Parameters: $w = 0.01, \alpha = 100, \mu = 0.1, \epsilon = 0.001, \gamma = 0.01, \kappa = 2, c = 1$, and $b = 10$. Initial conditions: $q = 0$ and $\beta = 0$.

co-evolution when varying the initial state of players over the entire simplex $\{(x_C, x_D, x_R) | x_\sigma \geq 0 \, \forall \sigma$ and $\sum_\sigma x_\sigma = 1\}$. The co-evolution reaches cooperative outcomes if there are sufficiently many conditional cooperators, *i.e.*, for sufficiently small $x_C$ and $x_D$, as initial conditions. The minimum fraction of conditional cooperators required to achieve a cooperative outcome when the initial state of players is a dimorphism of conditional and unconditional cooperators is smaller than that with a dimorphism of conditional cooperators and unconditional defectors.

# S5 Effects of the shape of the observation cost function on the outcome

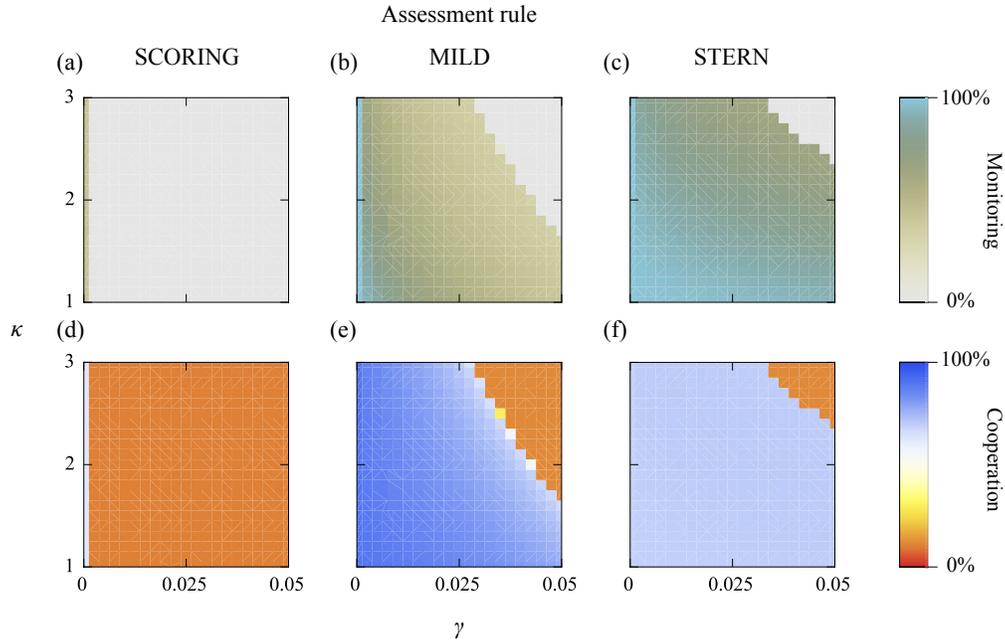

Figure S2: **Outcomes of co-evolution when varying parameters $\gamma$ and $\kappa$ in the observation cost function.** Panels show the equilibrium frequencies of **(a,b,c)** monitoring and **(d,e,f)** cooperation when varying with parameters $\gamma$ and $\kappa$ in the observation cost function (Eq. (S16)). **(a,d)** The SCORING rule. **(b,e)** The MILD rule. **(c,f)** The STERN rule. Parameters: $w = 0.01, \alpha = 100, \mu = 0.1, \epsilon = 0.001, c = 1$, and $b = 10$. Initial conditions: $x_C = 0, x_D = 0.5, x_R = 0.5, q = 0$, and $\beta = 0$.

Figure S2 shows the outcomes of co-evolution when varying the parameters $\gamma$ and $\kappa$ in $C(q)$. As expected, if $\gamma$ and $\kappa$ are large, the co-evolution cannot achieve high levels of monitoring and cooperation. Because both $\gamma$ and $\kappa$ have the same qualitative effect, we set $\kappa = 2$ throughout the main text and let $\gamma$ be the parameter representing the degree of observation cost.

# S6 Stable oscillations under the MILD rule

Under the MILD rule, we found a few parameter sets that produce stable oscillations in the co-evolutionary dynamics (shown in Fig. S3). Under the SCORING and STERN rules, we did not find any stable oscillations. The numerical simulations are conducted using the same parameter sets as those in Figs. 3 and 4 in the main text.

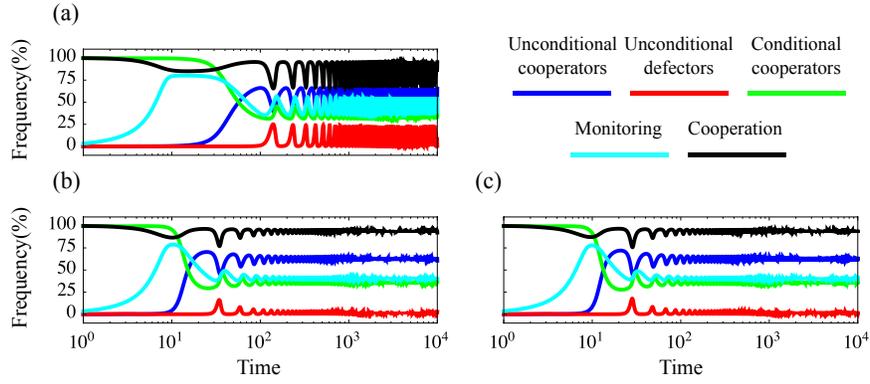

Figure S3: **Stable oscillations under the MILD rule.** Panels show the three cases of the stable oscillation under the MILD rule. Solid lines indicate the evolution of the frequencies of unconditional cooperators, unconditional defectors, and conditional cooperators (blue, red, and green curves, respectively), as well as those of monitoring (by monitors; cyan curve) and of cooperation (by unconditional or conditional cooperators; black curve) under the MILD rule. (a) When $w = 0.891251$ and $\alpha = 12.5893$. (b) When $w = 7.07946$ and $\alpha = 15.8489$. (c) When $w = 10$ and $\alpha = 15.8489$. Parameters: $\mu = 0.1, \epsilon = 0.001, \gamma = 0.01, \kappa = 2, c = 1$, and $b = 10$. Initial conditions: $q = 0, \beta = 0, x_C = 0, x_D = 1 - x_R$, and $x_R = 1$.

11


\end{document}